\documentclass[aps,prb,floatfix,twocolumn,showpacs,amsmath,amssymb]{revtex4}
\usepackage{graphicx}
\usepackage{dcolumn}
\usepackage{bm}
\usepackage{color}

\definecolor{light-gray}{gray}{0.85}

\begin{document}

\title{One-dimensional spin liquid, collinear, and spiral phases from uncoupled chains 
to the triangular lattice}
\author{Luca F. Tocchio,$^{1,2}$ Claudius Gros,$^{1}$
Roser Valent\'i,$^{1}$ and Federico Becca$^{2}$}
\affiliation{
$^{1}$Institute for Theoretical Physics, University of Frankfurt, 
       Max-von-Laue-Stra{\ss}e 1, D-60438 Frankfurt a.M., Germany \\
$^{2}$CNR-IOM-Democritos National Simulation Centre 
       and International School for Advanced Studies (SISSA), 
       Via Bonomea 265, I-34136, Trieste, Italy
            }

\date{\today} 

\begin{abstract}
We investigate the Hubbard model on the anisotropic triangular lattice with two hopping parameters 
$t$ and $t^\prime$ in different spatial directions, interpolating between decoupled chains ($t=0$) 
and the isotropic triangular lattice ($t=t^\prime$). Variational wave functions that include both 
Jastrow and backflow terms are used to compare spin-liquid and magnetic phases with different pitch 
vectors describing both collinear and coplanar (spiral) order. For relatively large values of the 
on-site interaction $U/t^\prime \gtrsim 10$ and substantial frustration, i.e., 
$0.3\lesssim t/t^\prime \lesssim 0.8$, the spin-liquid state is clearly favored over magnetic states. 
Spiral magnetic order is only stable in the vicinity of the isotropic point, while collinear order 
is obtained in a wide range of inter-chain hoppings from small to intermediate frustration. 
\end{abstract}


\maketitle

\section{Introduction}\label{sec:intro}

Since the pioneering work by Fazekas and Anderson 40 years ago,~\cite{fazekas1974} where magnetically 
disordered states (the so-called spin liquids) have been proposed as alternative ground states to 
standard ordered phases in magnetic systems, the field of frustrated magnetism has evolved as an 
important branch in condensed matter physics. The main motivation is that several unconventional 
features may arise when magnetism is suppressed at very low temperatures, like fractionalization 
of quantum numbers or topological degeneracy, to mention a few.

After four decades, our understanding on the subject is still rather incomplete: on the theoretical 
side, more effort is needed to clarify which are the microscopic models that may sustain spin-liquid 
ground states and which are suitable diagnostics to detect the presence of exotic properties; on the 
experimental side, it is important to synthesize and characterize new materials that may present both
strong electronic interactions and suitable magnetic frustration.

Organic charge-transfer salts, based on molecules like (BEDT-TTF)$_2$ or Pd(dmit)$_2$, represent 
important examples where the interplay between electronic itineracy, strong correlation, and 
geometrical frustration lead to various interesting phenomena.~\cite{definition} In these systems, 
the building blocks are given by extended molecular orbitals of dimerized molecules that are arranged 
in stacked triangular lattices. By varying the applied pressure and temperature, as well as the nature
of the cation associated to these materials, they may show metallic, superconducting, or insulating 
properties.~\cite{kanoda2011,powell2011} Whenever a single orbital for each dimer is considered,
the low-temperature behavior of these materials may be captured by a single-band Hubbard model on the 
anisotropic triangular lattice, with half-filled density, and relatively large on-site Coulomb 
repulsion.~\cite{powell2011,kandpal2009,nakamura2009}

Besides these organic systems, also Cs$_2$CuBr$_4$ and Cs$_2$CuCl$_4$ have a crystalline structure 
in which (magnetic) Copper atoms lie on weakly-coupled triangular lattices. In spite of being 
isostructural and isoelectronic, these two materials have completely different magnetic behavior. 
While the Br compound shows spiral magnetic ordering with well-defined magnon 
excitations,~\cite{ono2004} Cs$_2$CuCl$_4$ shows spin-liquid behavior over a broad temperature range 
with fractional spin excitations as revealed by inelastic neutron scattering 
experiments.~\cite{coldea2001} At very small temperatures, i.e., below $T_N=0.62 K$, the existence of 
a tiny inter-layer coupling stabilizes a true three-dimensional magnetic order. Also in this case, 
the low-energy properties of these materials may be captured by considering correlated electrons with 
highly reduced kinetic energy and strong Coulomb repulsion, possibly including different $3d$ orbitals 
of the Copper atoms in the Br compound.~\cite{foyevtsova2011} Cs$_2$CuBr$_4$ and Cs$_2$CuCl$_4$ are the end-member 
compounds of the family Cs$_2$CuCl$_{4-x}$Br$_x$ which shows a variety of magnetic properties when $x$ 
is changed.~\cite{cong2011} The distinct low-energy behaviors are generically attributed to the 
respective different degrees of frustration, which are determined in turn by the respective ratios 
between the inter-chain and intra-chain couplings in the underlying anisotropic triangular lattice.

\begin{figure}
\includegraphics[width=0.9\columnwidth]{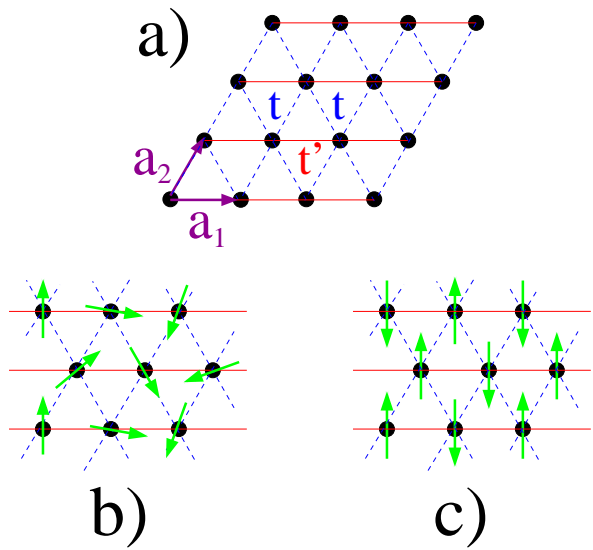}
\caption{\label{fig:lattice}
(Color on-line) Illustration of the anisotropic triangular lattice (a); solid and dashed lines denote 
hopping amplitudes $t^\prime$ and $t$, respectively. Spin patterns for the spiral with 
$\theta^\prime=2\theta$ (b) and for collinear (c) states.}
\end{figure}

Motivated by the rich phenomenology of these materials, we study the single-band Hubbard model on the 
anisotropic triangular lattice:
\begin{equation}\label{eq:hubbard}
{\cal H}=-\sum_{i,j,\sigma} t_{ij} 
c^\dagger_{i,\sigma} c_{j,\sigma}^{\phantom{\dagger}} + 
\textrm{h.c.} + U \sum_{i} n_{i,\uparrow} n_{i,\downarrow},
\end{equation}
where $c^\dagger_{i,\sigma} (c_{i,\sigma}^{\phantom{\dagger}})$ creates (destroys) an electron with 
spin $\sigma$ on site $i$ and $n_{i,\sigma}=c^\dagger_{i,\sigma}c_{i,\sigma}^{\phantom{\dagger}}$ 
is the electronic density; $U$ is the on-site Coulomb repulsion and $t_{ij}$ is the hopping amplitude,
including an intra-chain $t^\prime$, along ${\bf a}_1=(1,0)$, and an inter-chain $t$, along 
${\bf a}_2=(1/2,\sqrt{3}/2)$ and ${\bf a}_3={\bf a}_2-{\bf a}_1$, see Fig.~\ref{fig:lattice}(a). 

In the following, we consider clusters with periodic boundary conditions defined by the vectors 
${\bf T}_1=l {\bf a}_1$ and ${\bf T}_2=l {\bf a}_2$, in order to have $l \times l$ lattices 
with $L=l^2$ sites. The half-filled case, which is relevant for the aforementioned materials, is 
considered here.

One important difference between organic salts and Cs$_2$CuCl$_{4-x}$Br$_x$ is the degree of 
frustration given by the ratio between inter- and intra-chain couplings. In the case of organic salts,
$t^\prime/t<1$, thus implying a truly two-dimensional (frustrated) lattice geometry; by contrast, in 
the case of Cs$_2$CuCl$_{4-x}$Br$_x$, $t/t^\prime<1$, thus leading to a more one-dimensional 
(but frustrated) regime. When considering the strong-coupling limit, the comparison between neutron 
scattering experiments and theoretical calculations suggested that the ratio between inter- and 
intra-chain super-exchange couplings is $J/J^\prime \simeq 0.74$ and $0.33$ for 
Cs$_2$CuBr$_4$~\cite{ono2005} and Cs$_2$CuCl$_4$,~\cite{coldea2001} respectively, while the interlayer 
coupling $J_\perp$ in Cs$_2$CuCl$_4$ is estimated to be smaller than $10^{-2}J$. Similar values for 
$J$ and $J^\prime$ in both materials have been also found from the temperature dependence of the 
magnetic susceptibility.~\cite{zheng2005} Going back to the Hubbard model, these ratios are equivalent 
to $t/t^\prime \simeq 0.86$ and $0.57$, mostly in agreement with a density-functional theory (DFT) 
study of microscopic models for these compounds.~\cite{foyevtsova2011} By using electron spin resonance 
spectroscopy, very recent evaluations of the super-exchange couplings suggested smaller values of 
$J/J^\prime$ in both Br and Cl compounds, i.e., $J/J^\prime \simeq 0.4$ and $0.3$, 
respectively.~\cite{zvyagin2014} Despite quantitative differences in the super-exchange couplings 
(especially for the Br compound), all these observations indicate that Cs$_2$CuCl$_4$ is more 
one-dimensional than Cs$_2$CuBr$_4$. 

Here, we focus our attention on the region with $t/t^\prime<1$, completing our recent work that
analyzed the opposite case with $t^\prime/t<1$.~\cite{tocchio2009,tocchio2013,jacko2013}
As discussed above, the region  $t/t^\prime<1$ is suitable for describing Cs$_2$CuCl$_{4-x}$Br$_x$.  
Although these systems are Mott insulators with electrons almost fully localized on Copper atoms,
investigating the insulating behavior and a possible metal-to-insulating transition with the more 
general Hubbard model may unveil unforeseen new phenomena driven by the interplay of strong correlation
and frustration; these aspects, as described below, have been only marginally investigated in the past.

A substantial body of theoretical work has treated the anisotropic triangular lattice in the region 
$0 \le t/t^\prime \le 1$, however mostly in the infinite-$U$ limit, i.e., when only spin degrees of 
freedom are present. For quasi-one-dimensional lattices, it has been suggested that spins may display 
a collinear pattern, in sharp contrast with what is found in the classical limit,~\cite{weihong1999} 
where spins of neighboring chains form an angle of $90$ degrees. This claim has been originally 
proposed by a renormalization group approach,~\cite{starykh2007} and then supported by density-matrix 
renormalization group (DMRG) calculations on a three-leg spin tube.~\cite{chen2013} 

Incommensurate magnetism has been suggested by a functional renormalization group study to appear in 
a small region close to the isotropic point, with a spin-liquid state characterized by commensurate 
magnetic correlations being stabilized when moving towards the one-dimensional 
limit.~\cite{thomale2011} The existence of an essentially one-dimensional spin liquid phase in a wide 
regime of inter-chain couplings has been also obtained by the variational Monte Carlo approach, based 
upon Gutzwiller projected mean-field states.~\cite{yunoki2006,ogata2007,heidarian2009} 
A relatively extended one-dimensional disordered phase has been also suggested by exact 
diagonalizations~\cite{weng2006} and by spin-wave approaches.~\cite{hauke2011} while a one-dimensional 
dimer phase comes out from the SU(N) Hubbard-Heisenberg model, solved in the large-N 
limit.~\cite{mckenzie2001} DMRG calculations using pinned fields on the boundaries found incommensurate 
spin-spin correlations all the way from the isotropic point to the limit of decoupled 
chains,~\cite{weichselbaum2011} suggesting a more classical scenario with a pitch vector that 
continuously changes with $J/J^\prime$. In this respect, also Dzyaloshinskii-Moriya interactions have 
been proposed to be relevant in stabilizing spiral magnetic order with respect to the collinear 
one.~\cite{sorensen2011}

In contrast to the Heisenberg-model investigations, not so many studies have been performed directly 
on the Hubbard model on the anisotropic triangular lattice for $t/t^\prime<1$. We mention 
mean-field~\cite{powell2007} and variational Monte Carlo~\cite{watanabe2008} approaches, suggesting a 
magnetically ordered ground state for large on-site interactions. Instead, calculations within the 
variational cluster approximation suggested that an extended spin-liquid phase may be present close 
to the isotropic point $t=t^\prime$.~\cite{sahebsara2008,yamada2014,li2014} Finally, a possible 
spin-liquid region close to the metal-insulator transition has been suggested for the Hubbard model 
on the isotropic point, for example by the strong-coupling approach of Ref.~[\onlinecite{yang2010}].

\begin{figure}[t]
\includegraphics[width=1.0\columnwidth]{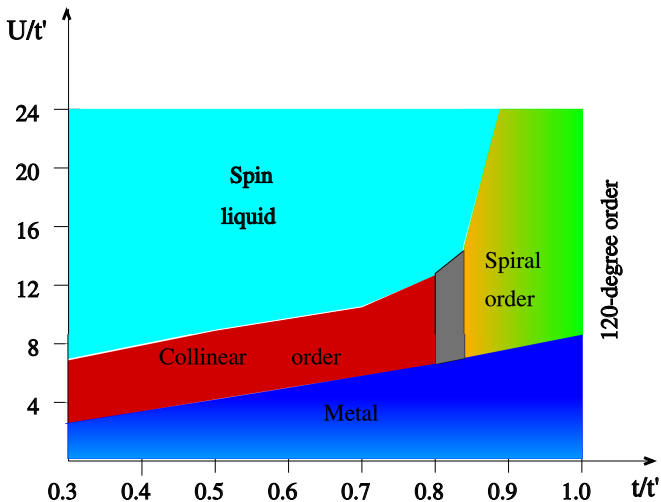}
\caption{\label{fig:diagram}
(Color online) Schematic phase diagram of the Hubbard model on the anisotropic triangular lattice 
with $t/t^\prime<1$, as obtained by variational Monte Carlo. In the spiral-order region, the optimal 
pitch angle ranges from $0.6\pi$ to $2\pi/3$, with the $120^{\circ}$ ordered state $\theta=2\pi/3$ 
being stable at the isotropic point. In the gray region $0.8 \lesssim t^\prime/t \lesssim 0.85$ 
the nature of the magnetic order cannot be reliably determined.} 
\end{figure}

Here, we consider improved variational wave functions, pursuing the approach used 
previously~\cite{tocchio2013} to study the case with $t^\prime/t<1$. In order to consider the relevant
phases that have been proposed for the Heisenberg model, we take into account correlated variational 
wave functions that may describe magnetic and spin-liquid states, as well as metallic or 
superconducting phases. In particular, concerning the magnetic case, we consider both wave functions 
having collinear magnetic correlations as well as states with spiral order, where the latter ones are 
constructed by starting from the states obtained at the Hartree-Fock 
level~\cite{krishamurthy1990,lacroix1993} and including, in a second step, many-body correlations. 
In this way, we are able to treat different magnetic and non-magnetic states on the same level and 
determine which state is stabilized for a given value of frustration $t/t^\prime$ and Coulomb 
repulsion $U$. We would like to mention that variational approaches may contain, as a matter of 
principle, a bias towards ordered states. However, very accurate results are obtained in a wide regime
of frustration when using, as in the present work, generalized Gutzwiller wave functions with 
long-range Jastrow terms and backflow corrections.~\cite{tocchio2008,tocchio2011} 

The paper is organized as follow: in section~\ref{sec:methods}, we discuss the form of the 
variational wave functions used in this work; in section~\ref{sec:results}, we present our
numerical calculations; finally, in section~\ref{sec:conc}, we draw the conclusions.

\section{Numerical methods}\label{sec:methods}

The variational wave functions that are used to draw the phase diagram as a function of $t/t^\prime$ 
and $U/t^\prime$ are given by:
\begin{equation}\label{eq:psi}
|\Psi\rangle = {\cal J}_s {\cal J}_d |\Phi_0\rangle,
\end{equation}
where ${\cal J}_s$ and ${\cal J}_d$ are conventional spin-spin and density-density Jastrow terms:
\begin{eqnarray}
{\cal J}_s &=& \exp \Big[ \frac{1}{2} \sum_{i,j} u_{i,j} S_i^z S_j^z \Big], 
\label{eq:jastrowspin} \\
{\cal J}_d &=& \exp \Big[ \frac{1}{2} \sum_{i,j} v_{i,j} n_i n_j \Big], 
\label{eq:jastrowdensity}
\end{eqnarray}
$u_{i,j}$ and the $v_{i,j}$ (that includes the on-site Gutzwiller term $v_{i,i}$) are pseudo-potentials
that can be optimized for every independent distance $|{\bf R}_i-{\bf R}_j|$ in order to minimize the 
variational energy and $S_i^z$ is the $z$-component of the spin operator on site $i$. $|\Phi_0\rangle$ 
is constructed starting from a generic mean-field Hamiltonian and then considering backflow 
correlations.~\cite{tocchio2008,tocchio2011} In particular, we will consider two possible mean-field 
Hamiltonians in order to describe either magnetic or paramagnetic states.  

\begin{figure*}[t]
\includegraphics[width=0.9\textwidth]{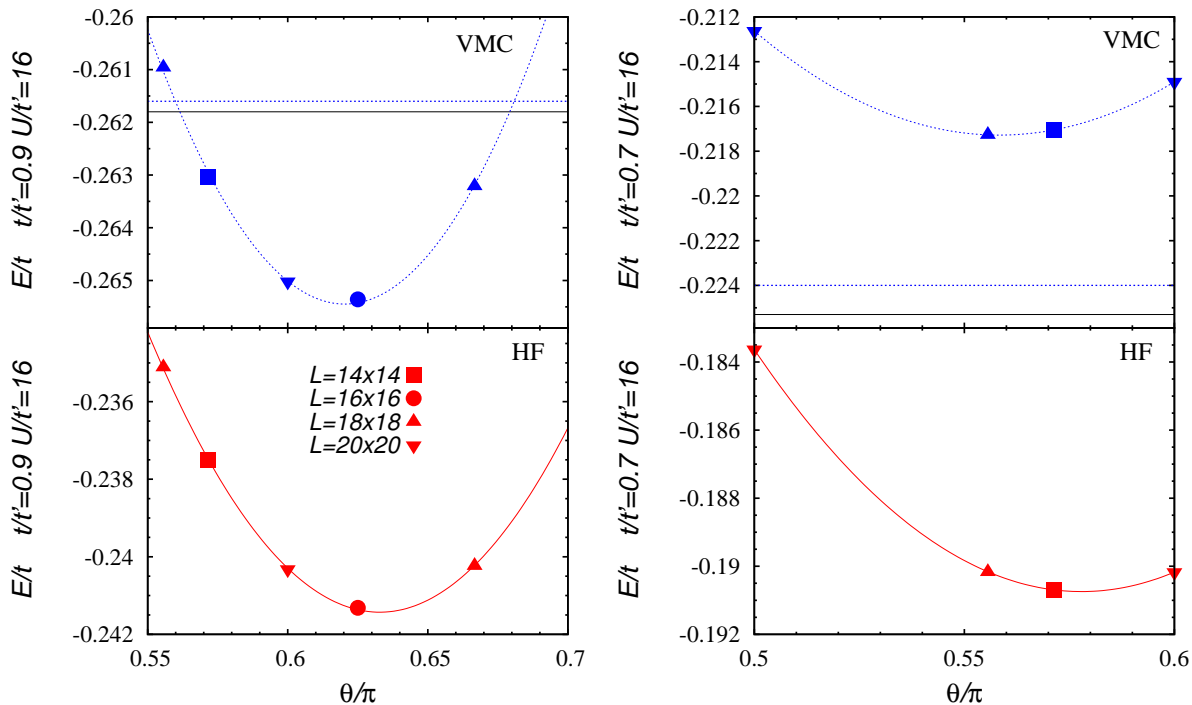}
\caption{\label{fig:energy_theta}
(Color on-line) Upper panels: Variational Monte Carlo energies of the spiral state as 
a function of the pitch angle $\theta$ for $U/t^\prime=16$ and $t/t^\prime=0.9$ (left panels) and 
$t/t^\prime=0.7$ (right panels). We show data for the largest lattice size to which each angle may
be accommodated. Lower panels: the same as in the upper panels, but for the Hartree-Fock calculations. 
In the upper panels, we also show for comparison the energy of the spin-liquid (horizontal solid black 
line) and of the collinear state (horizontal dotted blue line) on an $L=18\times 18$ lattice size.}
\end{figure*}

As far as the former ones are concerned, we perform the unrestricted Hartree-Fock decoupling of 
Eq.~(\ref{eq:hubbard}) as described in Ref.~[\onlinecite{tocchio2013}]. We then obtain the decoupled 
Hamiltonian:
\begin{eqnarray}\label{eq:Hspiral}
{\cal H}_{\textrm{AF}}&=&
-\sum_{i,j,\sigma} t_{ij} c^\dagger_{i,\sigma} c_{j,\sigma} + \textrm{h.c.} \\
&& + U \sum_i \left[ \langle n_{i,\downarrow} \rangle n_{i,\uparrow}
+ \langle n_{i,\uparrow} \rangle n_{i,\downarrow}\right] \nonumber \\
&& - U \sum_i \left[ 
\langle c^\dagger_{i,\uparrow} c_{i,\downarrow} \rangle 
c^\dagger_{i,\downarrow} c_{i,\uparrow} 
+ \langle c^\dagger_{i,\downarrow} c_{i,\uparrow} \rangle 
c^\dagger_{i,\uparrow} c_{i,\downarrow} \right] \nonumber \\
&& - U \sum_i \left[ \langle n_{i,\uparrow} \rangle 
\langle n_{i,\downarrow} \rangle - 
\langle c^\dagger_{i,\uparrow} c_{i,\downarrow} \rangle 
\langle c^\dagger_{i,\downarrow} c_{i,\uparrow} \rangle \right],
\nonumber
\end{eqnarray}
This Hamiltonian contains $4L$ independent mean-field parameters:
$\langle n_{i,\uparrow} \rangle$, $\langle n_{i,\downarrow} \rangle$,
$\langle c^\dagger_{i,\uparrow} c_{i,\downarrow} \rangle$, and
$\langle c^\dagger_{i,\downarrow} c_{i,\uparrow} \rangle$ for $i=1,...,L$
which have to be computed self-consistently. We impose the spin order to be coplanar in the $x{-}y$ 
plane, i.e., we look for solutions with 
$\langle n_{i,\uparrow} \rangle = \langle n_{i,\downarrow} \rangle$, 
which reduces the number of independent parameters to $3L$. The ground state  is then obtained
by diagonalizing the  mean-field Hamiltonian~(\ref{eq:Hspiral}).

In general, a regular spin pattern in the $x{-}y$ plane can be described by two angles $\theta$ and 
$\theta^\prime$, defining the relative orientation of two neighboring spins along ${\bf a}_2$ and 
${\bf a}_1$, respectively. Here, we focus on the insulating region of the phase diagram where, 
according to previous calculations for the Heisenberg model,~\cite{weihong1999} the optimal 
Hartree-Fock solutions display a spiral magnetic order, which may be parametrized through a single 
angle $\theta\in[\pi/2,2\pi/3]$, with $\theta^\prime=2\theta$, see Fig.~\ref{fig:lattice}(b). 
A pitch angle of $\theta=2\pi/3$ corresponds to the $120$-degree order, suitable for $t=t^\prime$, 
while $\theta=\pi/2$ corresponds to antiferromagnetic order along the chains with hopping $t^\prime$, 
the spins of neighboring chains forming an angle of $90$ degrees (appropriate for the limit $t \to 0$).
Besides this class of spiral states, we also consider states with collinear order, i.e., with 
$\theta^\prime=\pi$ and $\theta=0$ or $\pi$, see Fig.~\ref{fig:lattice}(c). 

We would like to mention that, in principle, also states with generic angles $\theta$ and 
$\theta^\prime$ would be possible; however, as shown in Sec.~\ref{sec:results}, we do not find any 
insulating region of the phase diagram where they give lower energies than the previous two magnetic 
states with $\theta^\prime=2\theta$ or with collinear patterns. Moreover, we do not find any evidence 
of a metallic phase with magnetic order.

Within this class of wave functions, both the spin and density Jastrow factors of 
Eqs.~(\ref{eq:jastrowspin}) and~(\ref{eq:jastrowdensity}) are important for the correct description 
of the low-energy properties. In particular, the spin term is fundamental to reproduce the spin-wave 
fluctuations above the mean-field state.~\cite{franjic1997,becca2000}

For the non-magnetic states (both metallic and insulating), we consider an uncorrelated wave function 
given by the ground state of a BCS Hamiltonian:~\cite{gros1988,zhang1988,gros1989,edegger2007}
\begin{equation}\label{eq:HBCS}
{\cal H}_{\rm{BCS}} = \sum_{k,\sigma} \xi_k 
c^\dagger_{k,\sigma} c_{k,\sigma} + \sum_{k} \Delta_k 
c^\dagger_{k,\uparrow} c^{\dagger}_{-k,\downarrow} + \textrm{h.c.},
\end{equation}
where the free-band dispersion $\xi_k$ and the pairing amplitude $\Delta_k$ are parametrized in the 
following way:
\begin{eqnarray}
\xi_k    &=& -2 \tilde{t}^\prime \cos ({\bf k} \cdot {\bf a}_1)
-2\tilde{t} [\cos ({\bf k} \cdot {\bf a}_2) \nonumber \\
         &+& \cos ({\bf k} \cdot {\bf a}_3)] -\mu, \\
\label{eq:epsilon}
\Delta_k &=& 2\Delta^\prime \cos ({\bf k} \cdot {\bf a}_1)
+2\Delta [\cos ({\bf k} \cdot {\bf a}_2) \nonumber \\
         &+& \cos ({\bf k} \cdot {\bf a}_3)], 
\label{eq:pairing}
\end{eqnarray}
with the effective hopping amplitude $\tilde{t}$, the effective chemical potential $\mu$, and the 
pairing fields $\Delta$ and $\Delta^\prime$ being variational parameters to be optimized 
($\tilde{t}^\prime$ gives the scale of energy of the mean-field Hamiltonian). In spite of the fact 
that we checked various symmetries for the pairing term, including both complex and non-translational 
invariant possibilities (see below), the best pairing function of Eq.~(\ref{eq:pairing}) is found to
have the $s+d_{xy}$ symmetry~\cite{powell2007} in all the range $t/t^\prime< 1$. In this regard,
the symmetry of the variational state that is optimized in presence of backflow and Jastrow terms 
agrees with previous calculations treating simpler wave functions.~\cite{powell2007,watanabe2008} 

\begin{figure}[t]
\includegraphics[width=\columnwidth]{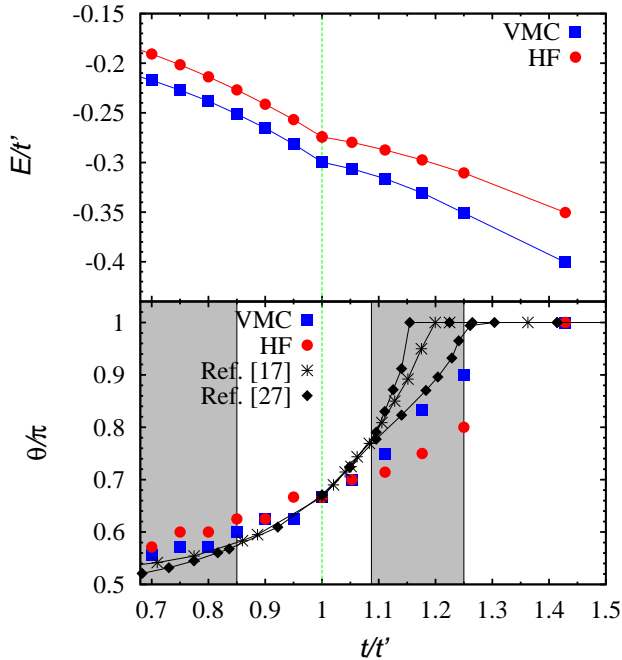}
\caption{\label{fig:HFvsVMC}
(Color online) Upper panel: Hartree-Fock (red circles) and variational (blue squares) energies of the 
optimal spiral state as a function of $t/t^\prime$ for $U/t^\prime=16$. Lower panel: the pitch angle 
$\theta$ (in unit of $\pi$) of the optimal spiral state as a function of $t/t^\prime$ for 
$U/t^\prime=16$. The data for $t>t^\prime$ are taken from Ref.~[\onlinecite{tocchio2013}]. The vertical
line denotes the isotropic point with $t=t^\prime$ and $\theta=2\pi/3$. The gray regions denote the 
values of $t/t^\prime$ where, according to variational Monte Carlo calculations at $U/t^\prime=16$, 
the spin-liquid state has an energy lower than the spiral one. Together with our Hartree-Fock and 
variational Monte Carlo calculations, we show for comparison the results for the Heisenberg model from 
Ref.~[\onlinecite{weihong1999}] and from Ref.~[\onlinecite{weichselbaum2011}]. In the latter case, two  
different sets of data are shown in the regime $t/t^\prime>1$, corresponding to $4n$ and $4n+2$ widths 
of the cylinders that are used in the DMRG calculations.}  
\end{figure}

In this case, we do not consider the spin Jastrow factor~(\ref{eq:jastrowspin}), since we do not want 
to break the spin SU(2) symmetry, retaining only density correlations~(\ref{eq:jastrowdensity}). 
The latter ones allow to obtain a non-magnetic Mott insulator for a sufficiently singular Jastrow 
factor $v_q \propto 1/q^2$ ($v_q$ being the Fourier transform of $v_{i,j}$), while $v_q \propto 1/q$ 
is found in a metallic/superconducting phase.~\cite{capello2005,capello2006} The wave function for the 
spin-liquid phase is the generalization to the Hubbard model of the fully-projected BCS state that has
been introduced by Anderson to describe the so-called resonating valence bond (RVB) state in the 
Heisenberg model.~\cite{anderson1987} 

We finally point out that the mean-field states obtained from Eqs.~(\ref{eq:Hspiral}) 
and~(\ref{eq:HBCS}) are supplemented by backflow terms, where each orbital that defines the
unprojected states is taken to depend upon the many-body configuration, in order to incorporate 
virtual hopping processes.~\cite{tocchio2008,tocchio2011} This procedure is a size-consistent 
and efficient way to improve the correlated wave functions on the lattice. Given the presence of 
Jastrow and backflow terms, the optimization of the variational wave functions by energy 
minimization and the calculations of all physical quantities must be performed by using quantum 
Monte Carlo techniques.~\cite{sorella2005}

\begin{figure}[t]
\includegraphics[width=0.9\columnwidth]{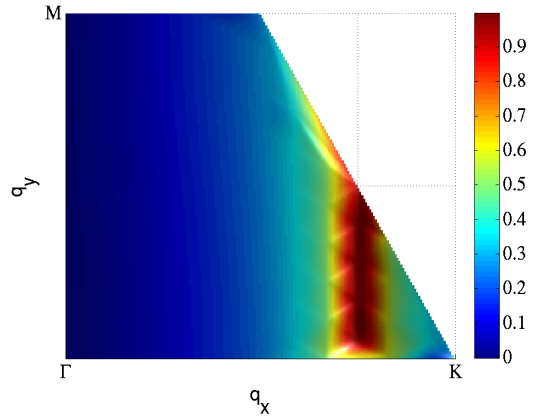}
\caption{\label{fig:Sq}
(Color online) Spin structure factor $S(q)$ computed with the spin-liquid wave function for 
$t/t^\prime=0.6$ and $U/t^\prime=16$. Results are reported in the first quarter of the Brillouin 
zone, ${\bf \Gamma}=(0,0)$, ${\bf K}=(4\pi/3,0)$, and ${\bf M}=(0,2\pi/\sqrt{3})$.}
\end{figure}

\section{Results}\label{sec:results}

The ground-state phase diagram as a function of the inter-chain hopping $t$ and of the Coulomb 
interaction $U$ is reported in Fig.~\ref{fig:diagram}, as obtained by comparing the energies of 
magnetic and non-magnetic wave functions. For small electron-electron interaction $U/t^\prime$ the 
ground state is metallic, the best state is obtained by starting from the mean-field \emph{ansatz} of 
Eq.~(\ref{eq:HBCS}) with very small pairing terms $\Delta$ and $\Delta^\prime$ and $v_q \propto 1/q$. 

The ground state turns out to be insulating for larger values of $U/t^\prime$, being either a spin 
liquid or magnetic. In the former case, the wave function is still described by the BCS \emph{ansatz}, 
with sizable pairing having an $s+d_{xy}$ symmetry and $v_q \propto 1/q^2$. In the latter case, the 
best variational state is constructed starting from the magnetic mean field~(\ref{eq:Hspiral}) that 
may have spiral or collinear order, with the collinear one being stabilized in a wide region for
$t/t^\prime \lesssim 0.8$. In the following, we describe in detail the properties of the different 
phases.

We point out that, on finite-size lattices with periodic-boundary conditions, only the set of pitch 
angles commensurate with the lattice size is accessible. Nonetheless, it is possible to reach a quite 
detailed understanding in the evolution of the wave vector describing ordered states. For $l \times l$
clusters, the allowed values are $\theta=2\pi n/l$, with $n$ being an integer. In this work, we will 
use lattice sizes ranging from $14 \times 14$ to $20 \times 20$.

\subsection{Magnetic states}

We find that in a remarkably wide region, i.e., for $t/t^\prime \lesssim 0.8$, the collinear state 
has a variational energy that is lower than the spiral wave function, while for 
$0.8 \lesssim t/t^\prime \le 1$ the spiral state is lower in energy. The optimal pitch angle is 
obtained by looking for the lowest energy state among a set of angles close to the value predicted by 
Hartree-Fock, see Fig.~\ref{fig:energy_theta}. We have also verified that states with non-trivial 
generic angles $\theta$ and $\theta^\prime$ do not provide a lower variational energy in the 
intermediate regime $t/t^\prime \simeq 0.8$, 

Collinear and spiral states are not connected continuously and a first-order phase transition could 
consequently occur between them; however, we cannot exclude that a very sharp crossover (with generic 
angles $\theta$ and $\theta^\prime$) appear in a narrow region around $t/t^\prime \simeq 0.8$.
The energies are too close to allow a reliable discrimination.

In Fig.~\ref{fig:HFvsVMC}, we present the evolution of the energy and of the pitch angle $\theta$ of 
the optimal spiral state as a function of $t/t^\prime$ for $U/t^\prime=16$, in the range where spiral 
order is relevant. The energy gain when Jastrow and backflow corrections are added to the mean-field 
wave function is non-negligible (about $0.03 t^\prime$ independently from $t$). Most importantly, the 
pitch angle is only slightly affected by the inclusion of electron correlations through the Jastrow 
and the backflow terms for $t/t^\prime<1$. This behavior is quite different from the case with 
$t^\prime/t <1$, which we also include in Fig.~\ref{fig:HFvsVMC}, where the correlation factors 
renormalize the mean-field angle more strongly.~\cite{tocchio2013} 

\subsection{Spin-liquid state} 

After optimization, the paramagnetic \emph{ansatz} has vanishingly small pairing terms for small 
Coulomb interactions, leading to a metallic phase for $U<U_c(t)$; as expected, we find that 
$U_c(t) \to 0$ for $t/t^\prime \to 0$, monotonically increasing up to the isotropic point $t=t^\prime$,
where $U_c(t=t^\prime) \simeq 8.5$.~\cite{kokalj2013} By increasing the electron interaction, the 
paramagnetic state turns insulating, because of the Jastrow factor that changes from $v_q \propto 1/q$ 
to $v_q \propto 1/q^2$. In this regime, the pairing terms $\Delta$ and $\Delta^\prime$ of 
Eq.~(\ref{eq:pairing}) become finite, with the $s+d_{xy}$ symmetry in all the range $t/t^\prime<1$; 
indeed, both complex or $d_{x^2-y^2}$ symmetries are never found to be optimal. These results are in 
good agreement with previous variational Monte Carlo calculations,~\cite{watanabe2008} where
however a small region with $d_{x^2-y^2}$ symmetry has been obtained close to the isotropic point. 
This discrepancy could be due to the more accurate treatment of electronic correlations in our 
approach. 

\begin{figure*}[t]
\includegraphics[width=0.9\textwidth]{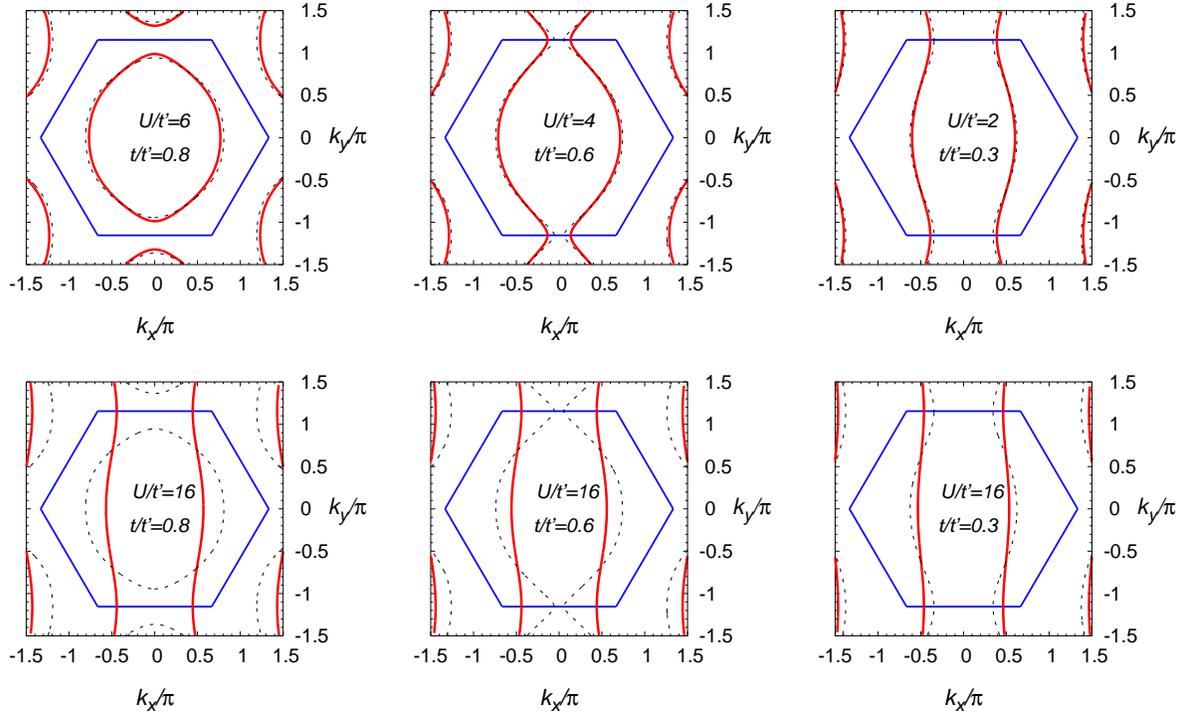}
\caption{\label{fig:FS}
(Color online) Upper panels: Fermi surface in the metallic phase (solid red lines) compared to the 
non-interacting results at $U=0$ (dashed black lines). Data are shown for three different values of 
$t/t^\prime$, close to the metal-insulator transition. Lower panels: Underlying Fermi surface in the 
spin-liquid region (solid red lines) compared to the non-interacting results (dashed black lines). 
Data are shown at $U/t^\prime=16$, for three different values of $t/t^\prime$. The Brillouin zone of 
the triangular lattice is marked by blue lines.}
\end{figure*}

\begin{figure*}[t]
\begin{minipage}{0.42\textwidth}
\includegraphics[width=\textwidth]{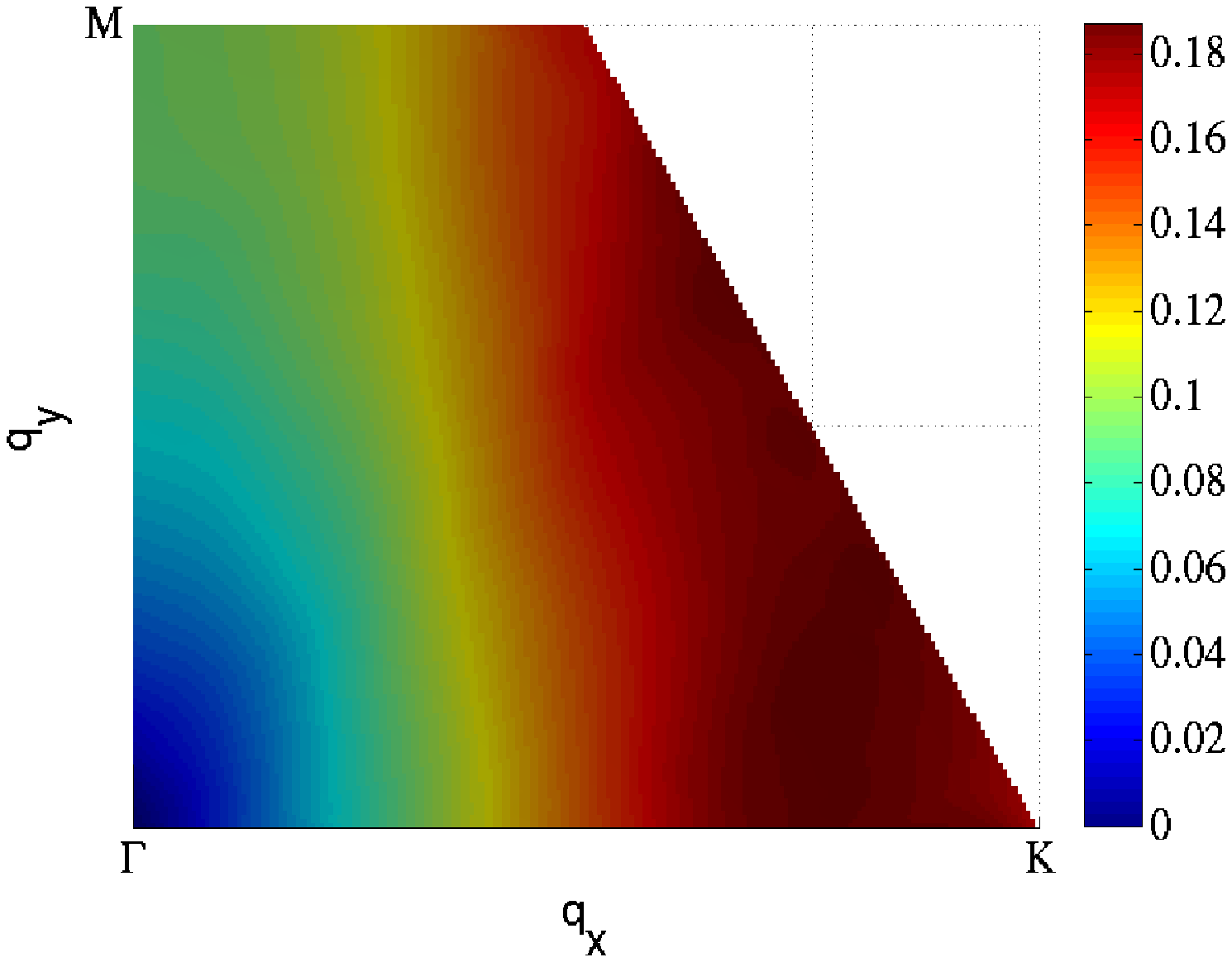}
\end{minipage}\hfill
\begin{minipage}{0.42\textwidth}
\includegraphics[width=\textwidth]{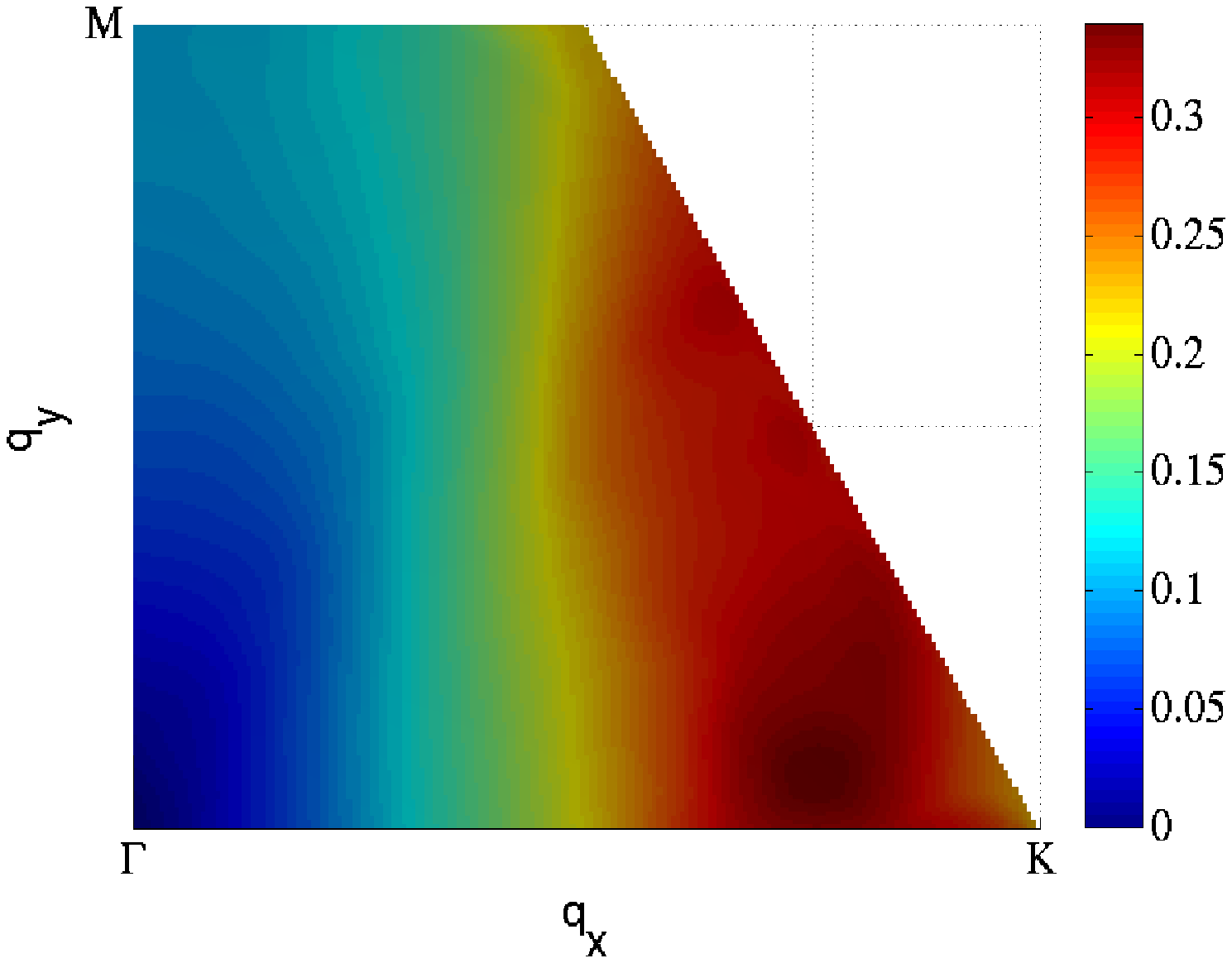}
\end{minipage}
\caption{\label{fig:Sq2}
(Color online) Spin-spin correlations $S(q)$ in the metallic phase for $U=0$ and $t/t^\prime=0.6$ 
(left panel) and for $U/t^\prime=4$ and $t/t^\prime=0.6$ (right panel). Results are reported in the 
first quarter of the Brillouin zone, ${\bf \Gamma}=(0,0)$, ${\bf K}=(4\pi/3,0)$, and 
${\bf M}=(0,2\pi/\sqrt{3})$.} 
\end{figure*}

Our result contrasts with variational calculations on the Heisenberg model, where two different 
spin liquids have been proposed, one with $s+d_{xy}$ symmetry for small frustrating regimes 
and another with a $2 \times 1$ unit cell close to the isotropic point.~\cite{yunoki2006} Instead, 
in the Hubbard model, the \emph{ansatz} with an extended $2\times 1$ unit cell in the mean-field state 
cannot be stabilized for finite electron-electron repulsion. We emphasize that, within the Heisenberg 
model, the energy difference between these two wave functions is very small (about $0.001J^\prime$) 
and, therefore, it is highly probable that in the presence of density fluctuations the more symmetric 
state that does not break translational symmetry is preferred.

We would like to point out that the spin-liquid state has strong one-dimensional features. Indeed, 
typical ratios for the pairings $\Delta$ and $\Delta^\prime$ are never larger than $0.1$ for a wide
range of inter-chain hoppings (i.e., for $0.1 \lesssim t/t^\prime \lesssim 0.9$), indicating that the 
pairing occurs essentially along the chains with hopping $t^\prime$. The relative sign of $\Delta$ 
and $\Delta^\prime$ changes around $t/t^\prime=0.5$, with $\Delta/\Delta^\prime > 0$ for 
$t/t^\prime < 0.5$ and $\Delta/\Delta^\prime < 0$ for $t/t^\prime \ge 0.5$; nevertheless, the 
mean-field spectrum is always gapless at four Dirac points, with the precise location of the Dirac 
points being dependent on the optimal value of the variational parameters. The one-dimensional nature 
of the spin-liquid phase is further confirmed by looking at the spin-spin correlations for the 
variational state:
\begin{equation}\label{eq:Sq}
S(q)=\frac{1}{L} \sum_{m,l} e^{i q (m-l)} \langle S_m^z S_l^z \rangle,
\end{equation}
where $S_m^z$ is the $z$-component of the spin operator on site $m$. Results are reported for 
$t/t^\prime=0.6$ and $U/t^\prime=16$, which is well inside the spin-liquid region, 
in Fig.~\ref{fig:Sq}. $S(q)$ exhibits an extended line of maxima with $q_x=\pi$, clearly indicating 
the one-dimensional character of the wave function. This fact is particularly remarkable given the 
relatively large inter-chain hopping. Moreover, these maxima do not diverge with $L$, showing that 
only short-range order is present. A strong one-dimensional character of the spin-liquid phase comes
out also from a mean-field study based upon Majorana fermions.~\cite{herfurth2013} A very anisotropic
magnon dispersion has been also proposed by a series expansion calculation.~\cite{fjaerestad2007}

It is important to stress the fact that the existence of the spin-liquid phase is due to the 
presence of a frustrating hopping $t$ between the chains. Indeed, if one-dimensional chains were 
coupled with an unfrustrated hopping, such to form an anisotropic square lattice with 
$t_x \ne t_y$, all the insulating phase would be immediately ordered with a commensurate N\'eel 
order. 

\subsection{Renormalization of the Fermi surface}

We present now some results on the properties of the Fermi surface in the metallic phase, as well as
the underlying Fermi surface in the insulating part of the phase diagram. In the upper panels of 
Fig.~\ref{fig:FS}, we report the Fermi surface $\xi_k=0$ in the metallic phase, close to the 
metal-insulator transition, for three values of the ratio $t/t^\prime$, compared to the non-interacting
case at $U=0$. Our results show only a weak renormalization of the Fermi surface, due to interaction.

However, we would like to remark that the spin-spin correlations $S(q)$, defined in Eq.~(\ref{eq:Sq}),
show a tendency toward one-dimensionality when increasing the Coulomb repulsion $U$ also in the 
metallic phase, see Fig.~\ref{fig:Sq2}. This result suggests that, even though the renormalization of
the Fermi surface is quite small in the metallic phase, this is sufficient to induce a noticeable
change of the spin-spin structure factor.

The concept of Fermi surface can be also generalized to systems that become gapped because of some 
symmetry breaking or electronic correlation, leading to the idea  of an underlying Fermi 
surface.~\cite{dzyaloshinskii2003} Within our variational approach, the underlying Fermi surface can 
be easily defined and corresponds, like in the metallic phase, to the locus of the highest occupied 
momenta $\xi_k=0$, when the pairing is set to zero.~\cite{tocchio2012} The lower panels of 
Fig.~\ref{fig:FS} show that the underlying Fermi surface in the spin-liquid region is strongly 
renormalized compared to the non-interacting case, leading to a quasi one-dimensional behavior. 
This result suggests that the collinear phase appearing at intermediate values of $U/t^\prime$ may be 
due to an instability of the spin-liquid region, and thus being a strong-coupling phenomenon.

\subsection{Phase diagram} 

\begin{figure}[t]
\includegraphics[width=\columnwidth]{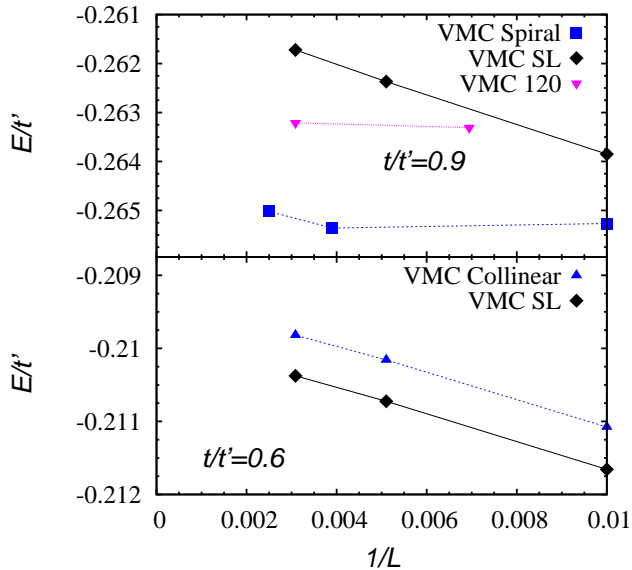}
\caption{\label{fig:spiral_SL}
(Color online) Upper panel: variational energies of the optimal spiral 
(blue squares) and spin-liquid (black diamonds) states as a function of 
the inverse system size $1/L$, for $U/t^\prime=16$ and $t/t^\prime=0.9$. The 
energy of the $120^{\circ}$ order with $\theta=2\pi/3$ is shown for comparison. 
Lower panel: variational energies of the collinear (blue triangles) and 
spin-liquid (black diamonds) states as a function of the inverse system size 
$1/L$, for $U/t^\prime=16$ and $t/t^\prime=0.6$.}
\end{figure}

By comparing the energies of magnetic and non-magnetic wave functions, we can obtain the 
zero-temperature phase diagram reported in Fig.~\ref{fig:diagram}. Few remarks are necessary to 
clarify some aspects of it. First of all, as mentioned above, given the fact that only a limited 
number of pitch angles are available on finite clusters, relatively large size effects are seen for 
spiral states. In various cases, the energy per site is not monotonic with $L$, showing that the 
true angle may be captured by some clusters but not by others. Even if these oscillations are never 
huge, i.e., of the order of $0.001 t^\prime$ for the sizes considered here, this fact makes it 
difficult to determine the precise boundary between spirals and other phases. 

Still, within our approach it is possible to obtain solid results in specific regimes. In particular, 
it is clear that a magnetically disordered phase is present for relatively large electron-electron
interactions and intermediate inter-chain hoppings. As an example, for $t/t^\prime=0.6$ and 
$U/t^\prime=16$, the lowest variational energy is achieved by the spin liquid state with the BCS 
\emph{ansatz}, see Fig.~\ref{fig:spiral_SL}. Here, the energies for the spin-liquid state are compared 
with the ones of the collinear state as a function of the cluster size, the spiral state being much 
higher in energy for all possible pitch angles (see Fig.~\ref{fig:energy_theta}). This is the typical 
outcome that appears for $U/t^\prime \gtrsim 10$ and $0.3 \lesssim t/t^\prime \lesssim 0.8$. 

When the ratio $t/t^\prime$ is reduced, the situation becomes less clear, since close to the 
one-dimensional limit both the collinear and the BCS states are very close in energy. For example, 
for $t/t^\prime=0.1$ we obtain $E/t^\prime=-0.3269(1)$ (for $U/t^\prime=8$) and $E/t^\prime=-0.1715(1)$ 
(for $U/t^\prime=16$) for both variational \emph{ans\"atze}; therefore, we cannot determine whether the 
spin-liquid phase persists down to $t \to 0$ or not. We should point out that since for $t/t^\prime=0.1$ 
the spin-liquid phase is not any more energetically favored at large values of $U/t^\prime$, our 
results do not exclude that the ground state of the Heisenberg model close to the one-dimensional limit
has (collinear) magnetic order, as predicted by Refs.~[\onlinecite{starykh2007,chen2013}]. It must be 
also emphasized that, in the one-dimensional limit both the BCS and the collinear state do not possess 
magnetic long-range order, since here the fluctuations generated by the Jastrow factor are sufficient 
to destroy the order present at the mean-field level.~\cite{franjic1997} On the other hand, whenever a 
coupling between chains is present, the magnetic order should not be destroyed by the Jastrow factor.

For $t/t^\prime \gtrsim 0.9$, the wave function with spiral order becomes competitive with both 
spin-liquid and collinear ones for  a non trivial angle $\theta/\pi$ that is about $0.6$. The trend 
is clear despite pronounced size effects, that are due to the relatively small number of angles that 
are commensurate with the lattice size, see Figs.~\ref{fig:energy_theta} and \ref{fig:spiral_SL}.

\section{Conclusions}\label{sec:conc}

In summary, we analyzed  the half-filled Hubbard model on the anisotropic lattice in the range of 
parameters $t/t^\prime<1$ by considering variational wave functions that include both Jastrow and 
backflow terms and are able to describe spin-liquid and magnetic states with different pitch vectors 
on the same footing. For large values of the interaction $U/t^\prime$ and  moderate to large 
inter-chain hoppings $0.3\lesssim t/t^\prime \lesssim 0.8$ a spin-liquid state with strong 
one-dimensional features and four Dirac points stabilizes with respect to any magnetic state 
considered here. However, close to the isotropic point, magnetic states with non trivial spiral order 
are more favorable. Interestingly, close to the metal-insulator transition, a collinear order 
stabilizes in a wide region of $t/t^\prime$, in sharp contrast with what is found in the classical 
limit. Our calculations are also relevant for the experimental results on the compounds Cs$_2$CuBr$_4$
and Cs$_2$CuCl$_4$, showing that in the strong coupling regime spiral magnetic order is stabilized 
close to the isotropic point, while a spin-liquid state emerges for smaller values of $t/t^\prime$.   

\acknowledgments

We thank H. Feldner for providing the Hartree-Fock code and A. Weichselbaum for the DMRG data shown in
Fig.~\ref{fig:HFvsVMC}. L.F.T., C.G. and R.V. acknowledge support by the German Science Foundation 
through the grant SFB/TRR49. F.B. and L.F.T. thank partial support from PRIN 2010-11. R.V. and F.B. 
acknowledge the KITP program {\it Frustrated Magnetism and Quantum Spin Liquids: From Theory and Models 
to Experiments} and the National Science Foundation under Grant No. NSF PHY11-25915.


\begin{thebibliography}{99}
\bibitem{definition} ET=BEDT-TTF stands for bis(ethylenedithio)-tetrathiafulvalene.
   dmit is 1,3-dithiole-2-thione-4,5-dithiolate.
\end{thebibliography}
\end{document}